\documentstyle[preprint,aps]{revtex}
\begin{document}
\draft
\title{Flavor changing neutral currents in $SU(3)\otimes U(1)$ models}
\author{ D. G\'omez Dumm$^a$, F. Pisano$^b$\thanks{Permanent Address:
Instituto de F\'\i sica Te\'orica, Universidade Estadual Paulista,
Rua Pamplona 145-900--S\~ao Paulo, SP, Brazil} and V. Pleitez$^c$}
\address{
$^a$ Laboratorio de F\'\i sica Te\'orica, \\
Departamento de F\'\i sica, U.N.L.P., \\
c.c. 67, 1900, La Plata, Argentina\\
$^b$ International Centre for Theoretical Physics, \\
Trieste, Italy\\
$^c$ Instituto de F\'\i sica Te\'orica,\\
Universidade Estadual Paulista,\\
Rua Pamplona 145\\ 01405-900-- S\~ao Paulo, S.P. \\ Brazil   }
\date{November 1993}

\maketitle

\begin{abstract}
We consider flavor changing neutral current effects coming from the $Z'$
exchange in 3-3-1 models. We show that the mass of this extra
neutral vector boson may be less than 2 TeV and discuss the problem of quark
family discrimination.
\end{abstract}


Recently it was proposed an electroweak model based on
the $SU(3)_L\otimes U(1)_N$ gauge symmetry~\cite{pp,pf}. The leptons are
treated democratically with the three generations transforming as
$({\bf3},0)$ but with one quark generation (it does not matter which
one) transforming differently from the other two. This condition arises
since the model must contain the same number of triplets and antitriplets
in order to be anomaly free. Hence, the number of generations
is related to the number of quark colors.

In Ref.~\cite{pp} the first generation is the one which transforms
differently from the second and the third ones. On the other hand, in
Ref.~\cite{pf} it was the third generation which was treated
differently. It was claimed that neutral currents could discriminate
between both choices of representation content~\cite{ng}. In fact, as we
will see below, this assessment is true only when further assumptions are
made about the quark mass matrices.

The GIM mechanism in several 3-3-1 models was consider in
Ref.~\cite{su3}. Here we turn back to the problem of flavor changing
neutral currents, showing in particular that the difference in
the choice of the quark representations is
less important than it was thought at first sight. The
lower bound for the mass of $Z'$ was overestimated in
Ref.~\cite{pp}.

We will use the notation of Ref.~\cite{pp}, but our results are
trivially written in the notation of Ref.~\cite{pf}. We also do not
consider the lepton sector here because there is no difference in it.

Let us start writing the quark representations: one of the
generations transforms as $({\bf3},2/3)$, denoting the second entry
$U(1)_N$ charges,
\begin{equation}
Q_{1L}=\left(
\begin{array}{c}
u_1 \\ d_1 \\ J
\end{array}
\right)
\label{q1}
\end{equation}
and the other two as $({\bf3}^*,-1/3)$:
\begin{equation}
Q_{2L}=\left(
\begin{array}{c}
j_1 \\ u_2 \\ d_2
\end{array}
\right),\qquad
Q_{3L}=\left(
\begin{array}{c}
j_2 \\ u_3 \\ d_3
\end{array}
\right).
\label{q23}
\end{equation}
The exotic quark $J$ has charge 5/3 and $j_i$, $i=1,2$ have both
charge $-4/3$. Eq.~(\ref{q1}) denotes the first or the third
generation. At this stage it does not matter this choice.

Denoting $U'=(u_1,u_2,u_3)^T$ and $D'=(d_1,d_2,d_3)^T$ the symmetry
eigenstates of charge 2/3 and $-1/3$ respectively, we can write in
this basis the neutral currents coupled with the extra neutral vector
boson $Z'$ \footnote{ The neutral currents coupled with
the $Z$ boson are
diagonal in the flavor space and we will not consider them here. }
\begin{equation}
{\cal L}_{Z'}=-\frac{g}{2\cos\theta_W}
(\bar U'_L\gamma^\mu Y^U_LU'_L+ \bar U'_R\gamma^\mu Y^U_RU'_R+
\bar D'_L\gamma^\mu Y^D_LD'_L+ \bar D'_R\gamma^\mu Y^D_RD'_R)Z'_\mu,
\label{nc}
\end{equation}
where
\begin{equation}
Y^U_L=Y^D_L=-\frac{1}{\sqrt{3}h(x)}\left(
\begin{array}{ccc}
1\,\, & \,\,       0\,\,   &\,\,    0 \\
0\,\, &\,\, -(1-2x)\,\, & \,\,   0  \\
0\,\, & \,\,       0\,\,   &\,\,  -(1-2x)
\end{array}
\right)
\label{yl}
\end{equation}
and
\begin{equation}
Y^U_R=-\frac{4x}{\sqrt{3}h(x)}\left(
\begin{array}{ccc}
1\,\, &\,\,        0 \,\,  & \,\,   0 \\
0\,\, &\,\,        1\,\,  & \,\,   0  \\
0\,\, &\,\,        0\,\,   &\,\,   1
\end{array}
\right), \quad Y^D_R=\frac{2x}{\sqrt{3}h(x)}\left(
\begin{array}{ccc}
1\,\, &\,\,        0 \,\,  & \,\,   0 \\
0\,\, &\,\,        1\,\,  & \,\,   0  \\
0\,\, &\,\,        0\,\,   &\,\,   1
\end{array}
\right).
\label{yr}
\end{equation}
Here we have defined $h(x)\equiv(1-4x)^{\frac{1}{2}}$, with
$x\equiv \sin^2\theta_W$.

In order to generate the quark masses, it is necessary to introduce the
following Higgs multiplets
\begin{equation}
\eta\sim({\bf3},0),\quad\rho\sim({\bf3},1),\quad\chi\sim({\bf3},-1),
\label{higgs}
\end{equation}
and a sextet $({\bf6},0)$ which is necessary in order to give masses to
the leptons \cite{fhpp}. Since this sextet does not couple to the quarks,
we need not to consider it here.

The Yukawa couplings for the charged 2/3 and $-1/3$ sector are
\begin{equation}
-{\cal L}_Y = \bar Q_{1L}(G_{1\alpha}U'_{\alpha
R}\eta+\tilde G_{1\alpha}D'_{\alpha R}\rho) +
\bar Q_{iL}(F_{i\alpha}U'_{\alpha R}\rho^*+\tilde F_{i\alpha}D'_{\alpha
R}\eta^*) + \mbox{ H.c.},
\label{yukawa}
\end{equation}
with $i=2,3$ and $\alpha=1,2,3$. $SU(3)$ indices have been suppressed
and $\eta^*,\rho^*$ denote the respective antitriplets~\cite{mpp}.

{}From Eq.~(\ref{yukawa}) it is straightforward to write the mass term
\begin{equation}
-{\cal L}_m=\bar U'_{\alpha
L}\Gamma^U_{\alpha\beta}U'_{\beta R}+
\bar D'_{\alpha L}\Gamma^D_{\alpha\beta} D'_{\beta R}+ \mbox{ H.c.},
\label{mass}
\end{equation}
where we have introduced the mass matrices
\begin{equation}
\Gamma^U=v_\eta
\left(\begin{array}{ccc}
G_{11} \,\, &\,\, G_{12}\,\, & G_{13} \\
F_{21}r & F_{22}r & F_{23}r \\
F_{31}r & F_{32}r & F_{33}r\end{array}\right),
\quad \Gamma^D=v_\rho\left(\begin{array}{ccc}
\tilde G_{11} \,\, & \,\,\tilde G_{12} \,\, & \,\,\tilde G_{13} \\
\tilde F_{21}/r  & \tilde F_{22}/r & \tilde F_{23}/r \\
\tilde F_{31}/r  & \tilde F_{32}/r & \tilde F_{33}/r
\end{array}
\right).
\label{1}
\end{equation}
Here, $v_\eta$ and $v_\rho$ represent the vacuum expectation values of
the neutral components of $\eta$ and $\rho$ respectively, and $r$ is the
ratio $v_\rho/v_\eta$. The mass matrices can be diagonalized by making the
biunitary transformations
\begin{mathletters}
\label{udmass}
\begin{equation}
U'_L=V^U_LU_L,\quad U'_R=V^U_RU_R,
\end{equation}
\begin{equation}
D'_L=V^D_LU_L,\quad D'_R=V^D_RD_R,
\end{equation}
\end{mathletters}
where the mass eigenstates are $U=(u,c,t)^T$ and $D=(d,s,b)^T$ if we
assume that the first generation is the one which transforms
differently, or $U=(t,u,c)^T$, $D=(b,d,s)^T$ if it is the third
generation which is treated in a different way. Both choices will
differ in the parameterization of the matrices in Eqs.\ (\ref{udmass}). We
will choose in the following the first alternative.

Using Eqs.\ (\ref{udmass}), we see from (\ref{nc}) that
\begin{mathletters}
\label{2}
\begin{equation}
Y^U_L\to V^{U\dagger}_L Y^U_L V^U_L,\quad
Y^U_R\to V^{U\dagger}_R Y^U_R V^U_R=Y^U_R.
\end{equation}
\begin{equation}
Y^D_L\to V^{D\dagger}_L Y^D_L V^D_L,\quad
Y^D_R\to V^{D\dagger}_R Y^D_R V^D_R=Y^D_R.
\end{equation}
\end{mathletters}
Notice that the right handed neutral currents remain diagonal, but not
the left-handed ones.

\hfill

Next, we want to consider possible constraints that arise from
experimental data in the $K^0-\bar K^0$, $D^0-\bar D^0$ and $B^0-\bar B^0$
systems. In particular, from Eqs.\ (\ref{nc}) and (\ref{2})
we can write the flavor-changing vertices $\bar d\gamma^\mu s$,
$\bar u\gamma^\mu c$ and $\bar d\gamma^\mu b$
\begin{mathletters}
\label{vuc}
\begin{equation}
{\cal L}_{ds} = \frac{g\cos\theta_W}{\sqrt{3}h(x)}[V^{D*}_{L11}
V^D_{L12}]\,\bar d_L\gamma^\mu s_LZ'_\mu + \mbox{ H.c.}
\end{equation}
\begin{equation}
{\cal L}_{uc} = \frac{g\cos\theta_W}{\sqrt{3}h(x)}[V^{U*}_{L11}
V^U_{L12}]\,\bar u_L\gamma^\mu c_LZ'_\mu + \mbox{ H.c.}
\end{equation}
\begin{equation}
{\cal L}_{db} = \frac{g\cos\theta_W}{\sqrt{3}h(x)}[V^{D*}_{L11}
V^U_{L13}]\,\bar d_L\gamma^\mu b_LZ'_\mu + \mbox{ H.c.}
\end{equation}
\end{mathletters}

Now from (\ref{vuc}) we obtain at first order in $G_F$ the effective
Lagrangian
\begin{equation}
{\cal L}^{eff}_{\Delta S=2}=\frac{G_F}{\sqrt{2}}
\frac{M_W^4}{M_Z^2M^2_{Z'}} \frac{4}{3h^2(x)}
(V_{L11}^{D\ast} V_{L12}^D)^2\,[\bar d_L\gamma^\mu s_L]^2,
\label{el1}
\end{equation}
together with the respective expressions for the $\Delta C\!=\!2$ and
$\Delta B\!=\!2$ operators. Moreover, defining $\Delta m_P=m_{P_1}-m_{P_2}$,
where $P_1$ and $P_2$ represent the neutral $K$, $D$ and $B$ mass eigenstates,
we obtain
\begin{equation}
\frac{\Delta m_P}{m_P}=\frac{G_F}{\sqrt{2}}
\frac{M_W^4}{M_Z^2M^2_{Z'}} \frac{8}{9h^2(x)} f_P^2 B_P
\mbox{Re}[(V_{L11}^{\ast} V_{L1j})^2]
\label{deltam}
\end{equation}
Here, $j\!=\!2$ for the $K_L-K_S$ and $D^0_1-D^0_2$ mass differences, and
$j\!=\!3$ in the $B^0-\bar B^0$ mixing case. The matrix $V$ has to be chosen
as the $V^U_L$ or the $V^D_L$ one depending on the quark type involved
in the corresponding mixing.

The experimental value for the mass differences are~\cite{pdg}
\begin{mathletters}
\label{values}
\begin{equation}
\Delta m_K = (3.522\pm0.016)\times 10^{-15}\,\mbox{GeV}
\end{equation}
\begin{equation}
\Delta m_D < 1.3\times 10^{-13}\,\mbox{GeV}
\end{equation}
\begin{equation}
\Delta m_B = (3.6\pm0.7)\times 10^{-13}\,\mbox{GeV}
\end{equation}
\end{mathletters}
Using $m_K\simeq 0.5$ GeV, $m_D\simeq 1.86$ GeV and $m_B\simeq 5.28$ GeV,
we obtain
\begin{equation}
\frac{\Delta m_K}{m_K}=7.04\times10^{-15},\quad
\frac{\Delta m_D}{m_D}\simeq 7.0\times 10^{-14},\quad
\frac{\Delta m_B}{m_B}\simeq 6.82\times 10^{-14}.
\label{values2}
\end{equation}
Assuming now that in all these cases the contribution (\ref{deltam}) to
$\Delta m_P/m_P$, coming from the $Z'$ exchange, is less than the
experimental values given in (\ref{values2}), we obtain the bounds
\begin{mathletters}
\label{ub}
\begin{equation}
M_{Z'} > 1.39\times 10^6\, [\mbox{Re}(V^{D*}_{L11}V^D_{L12})^2]^\frac{1}{2}
\,\mbox{GeV}
\end{equation}
\begin{equation}
M_{Z'} > 5.52\times 10^5\, [\mbox{Re}(V^{U*}_{L11}V^U_{L12})^2]^\frac{1}{2}
\,\mbox{GeV}
\end{equation}
\begin{equation}
M_{Z'} > 6.15\times 10^5\, [\mbox{Re}(V^{D*}_{L11}V^D_{L13})^2]^\frac{1}{2}
\,\mbox{GeV}.
\end{equation}
\end{mathletters}
We have used $f_K\approx 0.16$ GeV and $f_D\approx 0.2$ GeV,
$x\approx 0.2325$ and all other values from Ref.~\cite{pdg}. In order to get
the numerical estimations, we have
taken the ``bag constants'' $B_D$ \cite{mo} and $B_K$ equal to one, and used
the lattice calculation $\sqrt{B_{B_{d}}}f_{B_{d}}\simeq 0.22$ \cite{bb}.

The bounds in (\ref{ub}) have been found to depend on the matrix elements
$V^{U,D}_{L1j}$. As it is well known,
in the Standard Model both matrices $V_{L}^{U,D}$ appear only in the
combination
\begin{equation}
 V_{L}^{U \dagger} V_{L}^{D} = V_{CKM} ,
\label{vkm}
\end{equation}
and for this reason it is a usual convention to assume that $V_{CKM}
= V_{L}^{D}$, that is, $V_{L}^{U} \equiv {\bf 1}$. However, as we can see
from (\ref{vuc}), this is not the situation in the present case. Actually, in
the model of refs.~\cite{pp,pf} both matrices survive in different pieces of
the Lagrangian and it is too strong to set $V^U_{L} = {\bf 1}$. This assumption
would be also not stable against radiative corrections since all matrix
elements evolve with energy according to the renormalization group equation
\cite{rc}. Hence, the upper bounds
for $M_{Z'}$ depend on new parameters, which
have been introduced due to the special representation content of the model.

The complex numbers $V^{U,D}_{Lij}$ cannot be estimated from the present
experimental data. Indeed, the mixing matrices are only constrained by
the relation (\ref{vkm}). We see thus from (\ref{ub}) that it is possible to
have a neutral boson $Z'$ with a mass of about 1.5 TeV if
\begin{equation}
\mbox{Re}(V^{U,D\ast}_{L11} V^{U,D}_{L1j}) \sim 10^{-3} \;\; \forall\, j.
\label{vuvd}
\end{equation}

All the results, up to now, are common to both models of Refs.~\cite{pp,pf},
since the labels ``first'' or ``third'' generation are completely
meaningless. Indeed, it is possible to go from one choice to another just
using the transformations
\begin{equation}
V_L^{U,D} \rightarrow \left(
\begin{array}{ccc}
0\,\, & \,\,       0 \,\,  & \,\,   1 \\
0\,\, &\,\,        1\,\,  &\,\,    0  \\
1\,\, & \,\,       0\,\,   &\,\,   0
\end{array}
\right) V_L^{U,D}
\label{trans}
\end{equation}

\hfill

As it is pointed out above, we are not able to get any experimental
information about the matrices $V_L^{U,D}$, except for relation (\ref{vkm}).
However, the observed hierarchies among both fermion masses
and mixing angles have induced the physicists to propose different Ans\"atze
about the matrices $\Gamma^{U,D}$ defined in (\ref{1}). We will see that with
this assumptions the
equivalence between both $SU(3)_{L} \otimes U(1)_{Y}$ models would not exist
any more.

The hierarchy puzzle has given rise to many quark mass matrix models in the
last fifteen years. In order to show how strongly the $Z'$ bounds can be
affected, let us consider the simple scheme proposed by H.\ Fritzsch
\cite{fri}, in which the mixing matrix elements respect the hierarchy
\begin{equation}
V^{U,D}_{ij} \approx \left( \frac{m_{i}}{m_{j}} \right)^{\frac{1}{2}},
\;\; i < j
\label{fri}
\end{equation}
This can be obtained assuming mass matrices $\Gamma^{U,D}$ obeying
$\Gamma_{ij}\approx (m_{i} m_{j})^{\frac{1}{2}}$ \cite{chsh}.

Within this scenario, the experimental values for $\Delta m_{P}$ ($P = K,D,$
$ B_{d}, B_{s},$ ...) will imply respective bounds for $M_{Z'}$ depending
on which is the quark family treated in a different way from the other two.
The results are summarized in Table \ref{tabla}. The approximated values for
the quark masses have been taken from \cite{gl}.

In order to obtain numerical estimations, we have taken all the phases of
the matrix elements equal to zero. This cannot be true if the quark mixing
matrix is to be responsible for the observed CP violation in nature. The
inclusion of complex phases would conduce to a reduction in the bounds
appearing in Table \ref{tabla}. However, the hierarchical picture should not
be modified, unless we asked the phases $V_{L1j}$ in (\ref{deltam}) to
yield the particular result Re$[(V_{L11}^{\ast} V_{L1j})^2] \simeq 0$. In
this way, as emerges from the table, now the election of the third family is
favored if $Z'$ has to get a mass of ${\cal O}$(1 TeV), as it is claimed in
refs.~\cite{pf,ng}. The strongest bound for this election can arise
from the $B_s - \bar B_s$ mixing, whose suppression factor
within the Fritzsch model is $|V^D_{23} V^D_{33}| \approx 0.2.$

It is important to remark that the proposal of Fritzsch is certainly not
the unique one which conduce to results like those of Table
\ref{tabla}. Actually, the matrix texture
\begin{equation}
\Gamma = \left(
\begin{array}{ccc}
\alpha_{11} \,\, & \,\,\alpha_{12} & \,\,\alpha_{13} \\
\alpha_{21} \,\, & \,\,\beta_{22} & \,\,\beta_{23} \\
\alpha_{31} \,\, & \,\,\beta_{32} & \,\,\gamma
\end{array}
\right),
\label{jera}
\end{equation}
with $\alpha_{ij} \ll \beta_{ij} \ll \gamma$, is common to many quark
mass matrix Ans\"atze \cite{mod}. Although the numerical results in the
table might be modified, assuming the structure (\ref{jera}) is enough to
favor the differentiation of the third family in order to keep relatively
low bounds for $M_{Z'}$ \footnote{Exceptions of this type of models are those
which treat with ``democratic'' mass matrices \cite{demo}. Their mixing
angles are in general not small, conducing to high bounds on $M_{Z'}$.}.

\acknowledgments

We would like to thank the
Con\-se\-lho Na\-cio\-nal de De\-sen\-vol\-vi\-men\-to Cien\-t\'\i
\-fi\-co e Tec\-no\-l\'o\-gi\-co (CNPq) for total (FP) and partial
(VP) financial support. One of us (F.P.) would also like to thank to
the International Centre for Theoretical Physics, Trieste for the
kind hospitality.

\begin{table}
\caption{$M_{Z'}$ lower bounds within a Fritzsch-type Ansatz for the
quark mass matrices}
\begin{tabular}{cccc}
``Different'' family & $K - \bar K$ & $D - \bar D$ & $B_d -
\bar B_d$ \\
\hline
First (ref. [1]) & 315 TeV & 35 TeV & 25 TeV \\
Second & 315 TeV & 35 TeV & 25 TeV \\
Third (ref. [2]) & 10 TeV & 300 GeV & 25 TeV \\
\end{tabular}
\label{tabla}
\end{table}

\end{document}